# Multi-mode active control of friction, dynamic ratchets and actuators

Mikhail Popov[1,2,3,*], Qiang Li[1]

[1]Technische Universität Berlin, 10623 Berlin, Germany

[2]National Research Tomsk Polytechnic University, 634050 Tomsk, Russia

[3]National Research Tomsk State University, 634050 Tomsk Russia

[*]m@popov.name

**Abstract**

Active control of friction by ultrasonic vibration is a well-known effect with numerous technical applications ranging from press forming to micromechanical actuators. Reduction of friction is observed with vibration applied in any of the three possible directions (normal to the contact plane, in the direction of motion and in-plane transverse). In this work, we consider the multi-mode active control of sliding friction, where phase-shifted oscillations in two or more directions act at the same time. Our analysis is based on a macroscopic contact-mechanical model that was recently shown to be well-suited for describing dynamic frictional processes. For simplicity, we limit our analysis to a constant, load-independent normal and tangential stiffness and two superimposed phase-shifted harmonic oscillations, one of them being normal to the plane and the other in the direction of motion. As in previous works utilizing the present model, we assume a constant local coefficient of friction, with reduction of the observed force of friction arising entirely from the macroscopic dynamics of the system. Our numerical simulations show that the resulting law of friction is determined by just three dimensionless parameters. Depending on the values of these parameters, three qualitatively different types of behavior are observed: (a) symmetric velocity-dependence of the coefficient of friction (same for positive and negative velocities), (b) asymmetric dependence with respect to the sign of the velocity, but with zero force at zero velocity, and (c) asymmetric dependence with non-zero force at zero velocity. The latter two cases can be interpreted as a "dynamic ratchet" (b) and an actuator (c).

**Keywords**: Vibration, dual-mode, active control of friction, ratchet, actuator.

## 1 Introduction

Both static and sliding friction can be significantly reduced by vibration. This is a well-known phenomenon with numerous technical applications, e.g. in metal forming [1] or ultrasonic machining [2], as well as in stabilization of system dynamics, e.g. suppression of brake squeal. Since the 1950s the influence of vibration on friction has been studied experimentally [3] and various theoretical models have been proposed [4]. Reduction of friction has been observed both under the influence of oscillations in the contact plane and perpendicular to the plane [5]. Tolstoi [6] was one of the first to emphasize the importance of normal oscillations for the correct understanding of friction. Extensive studies have been carried out in [4][7] and in a series of dissertations [8]-[10]. Various configurations (oscillations in the sliding direction and perpendicular to the sliding direction in the contact plane; oscillations perpendicular to the contact plane (out-of-plane oscillations)) as well as a microscopic interpretation of the phenomenon are discussed in [11]. In a series of recent papers, it was shown experimentally that an important parameter in the problem of active control of friction is the contact stiffness [12][13]. This influence was analyzed in detail in [14] and [16] for the case of normal oscillations. In the present paper we consider the more general case of superimposed oscillations in normal and tangential directions ("dual-mode" active control of friction). We will show that this significantly changes the situation compared with "single-mode" control. In the case of dual-mode control a qualitatively different behavior can be observed, which we call "dynamic ratcheting".

## 2 Spring model and theoretical analysis

We consider an elastic body sliding on a flat plane with a constant velocity $v_0$, which is subjected to a superposition of external normal and tangential oscillations. It is assumed that Coulomb's law of friction with a constant local coefficient of friction $\mu_0$ is valid within the contact. Similarly to [14], we model the contact as a single linear spring with constant normal and tangential stiffness $k_z$ and $k_x$. This model corresponds to the contact of a flat-ended cylinder with a plane. The unstressed state in contact with the plane is chosen as the reference state. The vertical and horizontal displacements of the upper point of the spring are denoted as $u_z$ and $u_x$, and the horizontal displacement of the lower point as $u_{x,c}$. The upper point experiences a forced oscillation according to

$$u_z = u_{z,0} - \Delta u_z \cos \omega t \qquad (1)$$

in the vertical direction, and a composition of translation with constant velocity and a harmonic oscillation with the same frequency $\omega$ and a phase difference $\varphi$

$$u_x = v_0 t + \Delta u_x \cos(\omega t + \varphi) \qquad (2)$$

in the horizontal direction (Fig. 1), where $u_{z,0}$ corresponds to the average indentation depth and $\Delta u_z$ and $\Delta u_x$ are the amplitudes of oscillation. We assume that the spring is always in contact with the plane, i.e. $\Delta u_z < u_{z,0}$. The main difference compared to [14] is the presence of the tangential oscillation $\Delta u_x \cos(\omega t + \varphi)$ in (2). This change results in qualitatively changed, ratchet-like or actuator-like behavior.

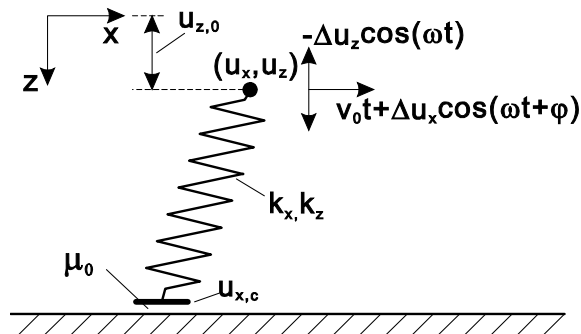

**Fig. 1** A single-spring model of frictional contact under superimposed oscillation in normal and tangential directions.

### 2.1 Critical velocity

One of the characteristics of reduction of friction by vibration is the existence of a critical sliding velocity, above which the reduction is no longer possible and the average coefficient of friction is equal to its local value $\mu_0$. In [14] and [16], it was shown that this critical velocity can be calculated from the contact-mechanical model for fairly general system configurations. These calculations were based on the observation that reduction of friction must be due to intermittent stick states during sliding, since the tangential force is *less* than the ordinary sliding friction force only during stick. The critical sliding velocity is thus determined by considering the point where stick becomes impossible and the system transitions into continuous sliding. In [15] this analysis was further generalized for the case of simple viscoelastic contacts. In the following we will calculate the critical velocity of controllability for dual mode control of friction.

As described above, the upper point of spring is forced to move according to Eqs. (1) and (2). However, the movement of lower point is unknown and can be either in stick or slip state. The normal and tangential force of the spring are given by

$$f_z = k_z u_z = k_z \left( u_{z,0} - \Delta u_z \cos \omega t \right), \tag{3}$$

$$f_x = k_x \left( u_x - u_{x,c} \right) = k_x \left( v_0 t + \Delta u_x \cos(\omega t + \varphi) - u_{x,c} \right). \tag{4}$$

If the tangential force of the spring is smaller than the normal force multiplied by the coefficient of friction, $f_x < \mu_0 f_z$, the lower point will be in a stick state. Otherwise, it will slip relative to the plane and in this case $f_x = \mu_0 f_z$:

$$k_x \left( v_0 t + \Delta u_x \cos(\omega t + \varphi) - u_{x,c} \right) = \mu_0 k_z \left( u_{z,0} - \Delta u_z \cos \omega t \right). \tag{5}$$

To find the critical velocity for continuous sliding, we assume that (5) is fulfilled at all times. Derivation of (5) with respect to time gives

$$\dot{u}_{x,c} = v_0 - \omega \Delta u_x \sin(\omega t + \varphi) - \mathrm{sgn}(v_0) \cdot \mu_0 \left( k_z / k_x \right) \omega \Delta u_z \sin \omega t \tag{6}$$

or

$$\dot{u}_{x,c} = v_0 - \omega \left[ \left( \Delta u_x \cos \varphi + \mathrm{sgn}(v_0) \cdot \mu_0 \left( k_z / k_x \right) \Delta u_z \right) \sin \omega t + \left( \Delta u_x \sin \varphi \right) \cos \omega t \right]. \tag{7}$$

The lower point will slide continuously in one direction if its velocity does not change sign or turn to zero at any point. By looking at Eq. (7), we see that this is the case if the constant part of the velocity is larger than the amplitude of the oscillating part. The latter will also be equal to the sought critical velocity:

$$\begin{aligned} v_0 > v_c &= \omega \sqrt{\left( \Delta u_x \cos \varphi + \mathrm{sgn}(v_0) \cdot \mu_0 \left( k_z / k_x \right) \Delta u_z \right)^2 + \left( \Delta u_x \sin \varphi \right)^2} \\ &= \omega \sqrt{\left( \Delta u_x \right)^2 + 2 \mu_0 \left( k_z / k_x \right) \Delta u_x \Delta u_z \mathrm{sgn}(v_0) \cdot \cos \varphi + \left( \mu_0 \left( k_z / k_x \right) \Delta u_z \right)^2} \end{aligned}. \tag{8}$$

For $v_0 > 0$, the critical velocity is maximal

$$v_{c.\max} = \omega \left( \Delta u_x + \mu_0 \left( k_z / k_x \right) \Delta u_z \right) \tag{9}$$

when the phase difference is zero, and reaches its minimal value

$$v_{c.\min} = \omega \left| \Delta u_x - \mu_0 \left( k_z / k_x \right) \Delta u_z \right| \tag{10}$$

at $\varphi = \pi$. For $v_0 < 0$, the corresponding phases are swapped.

It is interesting to note that for positive $v_0$, if $\varphi = \pi$ and $\Delta u_x = \mu_0 \left( k_z / k_x \right) \Delta u_z$, the critical velocity is equal to zero, which means that the coefficient of friction remains unchanged at any velocity $v_0$. Since only the ratio of the stiffness appears here, this result should be independent of the indenter shape at least for small oscillation amplitudes.

## 2.2 Static friction

Another distinctive feature of the influence of vibration on friction is the static force of friction at zero sliding velocity. This can also be calculated analytically in many cases. The static force of friction is the largest force that does not result in slip. For this to be true, the lower point of the spring must not move from its initial position: $u_{x,c} = 0$ and the tangential force on the spring must remain less than the force of friction at all times:

$$|f_x| < \mu_0 |f_z| \tag{11}$$

or

$$|k_x (u_{x,0} + \Delta u_x \cos(\omega t + \varphi))| < \mu_0 |k_z (u_{z,0} - \Delta u_z \cos \omega t)| \tag{12}$$

Here $u_{x,0}$ is the equilibrium tangential displacement of the spring, around which $u_x(t)$ oscillates. Remember that we assumed that the spring is always in contact with the substrate and the normal force thus always nonnegative. For this reason we can drop the modulus on the right hand side of (12). For the following analysis we will also drop the modulus on the left hand side. This makes our calculations less than perfectly rigorous and numerical simulations confirm that this results in incorrect static force for some values of $\varphi$. Nonetheless, this assumption seems to be valid in *most* cases, which is why we present the following, admittedly incomplete, analysis.

With the above assumption, we solve for $u_{x,0}$ and obtain:

$$\begin{aligned} u_{x,0} &< \mu_0 \frac{k_z}{k_x} u_{z,0} - \mu_0 \frac{k_z}{k_x} \Delta u_z \cos \omega t - \Delta u_x \cos(\omega t + \varphi) = \\ &\mu_0 \frac{k_z}{k_x} u_{z,0} - \cos \omega t \left( \mu_0 \frac{k_z}{k_x} \Delta u_z + \Delta u_x \cos \varphi \right) + \Delta u_x \sin \omega t \sin \varphi \end{aligned} \tag{13}$$

Thus, the stick condition is satisfied if

$$\begin{aligned} u_{x,0} &< \mu_0 \frac{k_z}{k_x} u_{z,0} - \sqrt{\left( \mu_0 \frac{k_z}{k_x} \Delta u_z + \Delta u_x \cos \varphi \right)^2 + \Delta u_x^2 \sin^2 \varphi} = \\ &\mu_0 \frac{k_z}{k_x} u_{z,0} - \sqrt{\left( \mu_0 \frac{k_z}{k_x} \Delta u_z \right)^2 + 2 \mu_0 \frac{k_z}{k_x} \Delta u_z \Delta u_x \cdot \cos \varphi + \Delta u_x^2} \end{aligned} \tag{14}$$

The maximum equilibrium displacement is maximized when $\varphi = \pi$:

$$u_{x,0} < \mu_0 \frac{k_z}{k_x} u_{z,0} - \left| \mu_0 \frac{k_z}{k_x} \Delta u_z - \Delta u_x \right| \tag{15}$$

and is minimized at $\varphi = 0$

$$u_{x,0} < \mu_0 \frac{k_z}{k_x} u_{z,0} - \left| \mu_0 \frac{k_z}{k_x} \Delta u_z + \Delta u_x \right|. \tag{16}$$

The static friction force is obtained by multiplying $u_{x,0}$ with the tangential stiffness:

$$F_s = k_x \left( \mu_0 \frac{k_z}{k_x} u_{z,0} - \sqrt{\left( \mu_0 \frac{k_z}{k_x} \Delta u_z \right)^2 + 2 \mu_0 \frac{k_z}{k_x} \Delta u_z \Delta u_x \cdot \cos \varphi + \Delta u_x^2} \right). \tag{17}$$

Dividing by the average normal force finally gives us the static coefficient of friction:

$$\frac{\mu_s}{\mu_0} = \frac{F_s}{k_z u_{z,0}} = 1 - \sqrt{\left( \frac{\Delta u_z}{u_{z,0}} \right)^2 + \frac{2}{\mu_0} \frac{k_x}{k_z} \frac{\Delta u_z \Delta u_x}{u_{z,0}^2} \cdot \cos \varphi + \left( \frac{k_x \Delta u_x}{\mu_0 k_z u_{z,0}} \right)^2}. \tag{18}$$

Once again, for $v_0 > 0$, if $\varphi = \pi$ and $\dfrac{\Delta u_z}{u_{z,0}} = \dfrac{k_x \Delta u_x}{\mu_0 k_z u_{z,0}}$ or $\dfrac{k_x \Delta u_x}{\mu_0 k_z \Delta u_z} = 1$, then $\mu_s = \mu_0$.

Note that the above is valid if the force is applied in the positive *x*-direction. If the force acts in the opposite direction without changing the polarity of oscillation, then Eq. (14) becomes

$$|u_{x,0}| < \mu_0 \frac{k_z}{k_x} u_{z,0} - \sqrt{\left(\mu_0 \frac{k_z}{k_x} \Delta u_z\right)^2 - 2\mu_0 \frac{k_z}{k_x} \Delta u_z \Delta u_x \cdot \cos\varphi + \Delta u_x^2} \; , \qquad (19)$$

with equivalent changes to Eqs. (17) and (18). The maximum displacement (15) is reached at $\varphi = 0$ and the minimum displacement (16) at $\varphi = \pi$.

## 3 Numerical simulation

While the critical velocity and the static coefficient of friction can be calculated in closed form in our model, the overall dependence of the coefficient of friction on sliding velocity cannot. The detailed dependences were therefore obtained numerically (by explicit integration). The macroscopic coefficient of friction was determined as the average value of tangential force divided by the average normal force in one oscillation period:

$$\mu = \langle f_x \rangle / \langle f_z \rangle, \qquad (20)$$

where the normal and tangential force are calculated according to Eqs. (3) and (4) in every time step. It should be pointed out that by this definition $\mu$ has the same sign as the sliding velocity, since we do not take the modulus. Also, as will be shown in a moment, in some cases $\mu$ can have a sign *opposite* to that of the velocity, which allows the system to function as a vibrational motor or actuator.

During integration, the coordinate of the contact point is updated whenever the tangential force exceeds the current maximum frictional force. In such a case the contact point is moved such that the spring is shortened and the two forces match. It should also be noted that results are only presented for the *steady state*. It may take several cycles of oscillation for the tangential force or stress to reach its equilibrium value. Especially at low velocities this may take a relatively long time. Any such kinetic effects are not subject of this study.

### 3.1 Single-mode oscillation

We first present results for single-mode control with either purely normal or purely tangential oscillation. Fig. 2 and Fig. 3 show the dependences of the macroscopic coefficient of friction on sliding velocity for these two special cases. In Fig. 2, only normal oscillation is applied ($\Delta u_x = 0$). This case was already discussed in detail in paper [14], where the following numerical approximation (accurate to within 1%) for the velocity-dependence of the coefficient of friction was obtained:

$$\frac{\mu}{\mu_0} \approx 1 - \frac{\Delta u_z}{u_{z,0}} \left[ \frac{3}{4}\left(\frac{v_0}{v_c} - 1\right)^2 + \frac{1}{4}\left(\frac{v_0}{v_c} - 1\right)^4 \right]. \qquad (21)$$

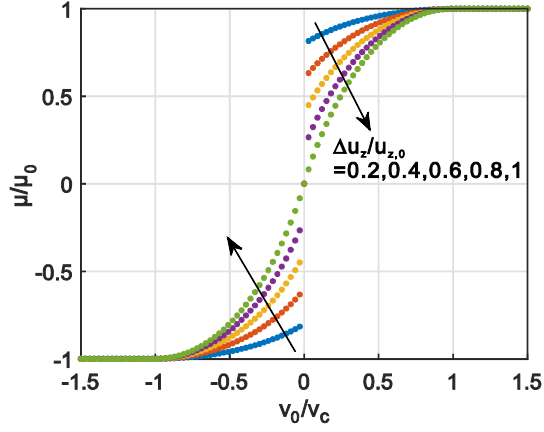

**Fig. 2** Dependence of the macroscopic coefficient of friction on sliding velocity for the case of purely normal oscillation ($\Delta u_x = 0$), for details see [14].

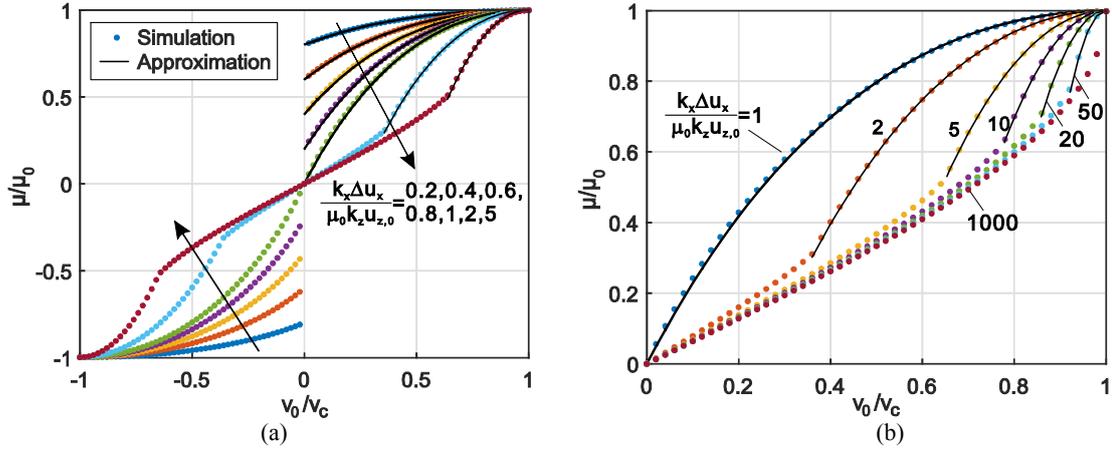

(a)          (b)

**Fig. 3** Dependence of macroscopic coefficient of friction on sliding velocity for the case of purely tangential oscillation ($\Delta u_z = 0$): (a) for small values of $k_x \Delta u_x / (\mu_0 k_z u_{z,0})$; (b) for large values of $k_x \Delta u_x / (\mu_0 k_z u_{z,0})$. The solid lines correspond to the approximation (24) and the dots represent numerical results.

Let us now turn to purely tangential oscillation. This case has been considered in the literature for very small system stiffness [5], but it has never been completely analyzed for a stiff system and a soft contact. Results for this case (with $\Delta u_z = 0$) are presented in Fig. 3. In this case, the critical value of sliding velocity, according to Eq. (8), reduces to

$$v_c = \omega \Delta u_x, \tag{22}$$

which is simply the velocity amplitude of tangential oscillation. Numerical simulations show that the coefficient of friction in this case is a function of only two dimensionless parameters: $v_0 / v_c$ and $\dfrac{k_x \Delta u_x}{\mu_0 k_x u_{z,0}}$:

$$\frac{\mu}{\mu_0} = f\left(\frac{v_0}{v_c}, \frac{k_x \Delta u_x}{\mu_0 k_z u_{z,0}}\right), \text{ for } \Delta u_z = 0. \tag{23}$$

Furthermore, for small values of the parameter $\dfrac{k_x \Delta u_x}{\mu_0 k_x u_{z,0}} \leq 1$, the curves are exactly the same as in the case of purely normal oscillation (Fig. 2), and can therefore be described with a very similar approximation:

$$\frac{\mu}{\mu_0} \approx 1 - \frac{k_x \Delta u_x}{\mu_0 k_z u_{z,0}} \left[ \frac{3}{4}\left(\frac{v_0}{\omega \Delta u_x} - 1\right)^2 + \frac{1}{4}\left(\frac{v_0}{\omega \Delta u_x} - 1\right)^4 \right], \quad \text{for } \frac{k_x \Delta u_x}{\mu_0 k_z u_{z,0}} \leq 1. \tag{24}$$

The results of numerical simulation and the approximation (24) are compared in Fig. 3a. For values of $\frac{k_x \Delta u_x}{\mu_0 k_x u_{z,0}} > 1$, the dependence still coincides with Eq. (24) at large velocities, but at low velocities or very large values of $\frac{k_x \Delta u_x}{\mu_0 k_x u_{z,0}}$, there are significant deviations. This is shown in more detail in Fig. 3b. For very large values of the dimensionless parameter it can be seen that the dependence becomes roughly linear.

### 3.2 General case: bi-modal oscillation

In the two cases considered above, the dependences are symmetric for positive and negative sliding velocity. In the following we will consider more general cases, which produce some interesting phenomena. First, both dimensional analysis and numerical results show that the coefficient of friction can be presented in the most general case as a function of four dimensionless parameters:

$$\frac{\mu}{\mu_0} = f\left(\frac{v_0}{v_c}, \frac{\Delta u_z}{u_{z,0}}, \frac{k_x \Delta u_x}{\mu_0 k_z u_{z,0}}, \varphi\right) \tag{25}$$

or

$$\frac{\mu}{\mu_0} = f\left(\frac{v_0}{v_c}, \frac{\Delta u_z}{u_{z,0}}, \frac{k_x \Delta u_x}{\mu_0 k_z \Delta u_z}, \varphi\right) \tag{26}$$

if this choice is more convenient. It is clear that Eq. (23) is just the limiting case of (25) for $\Delta u_z = 0$.

Qualitatively different behaviors for the case of zero phase shift are shown in Fig. 4, where the normalized coordinates $\mu/\mu_0$ and $v_0/v_c$ are used. The dependence of the coefficient of friction on sliding velocity becomes asymmetric and sometimes qualitatively different for the positive and negative sliding directions. At small positive velocities (still with $\varphi = 0$), a negative coefficient of friction (opposite frictional force) is observed, especially if the amplitude of normal oscillation is large.

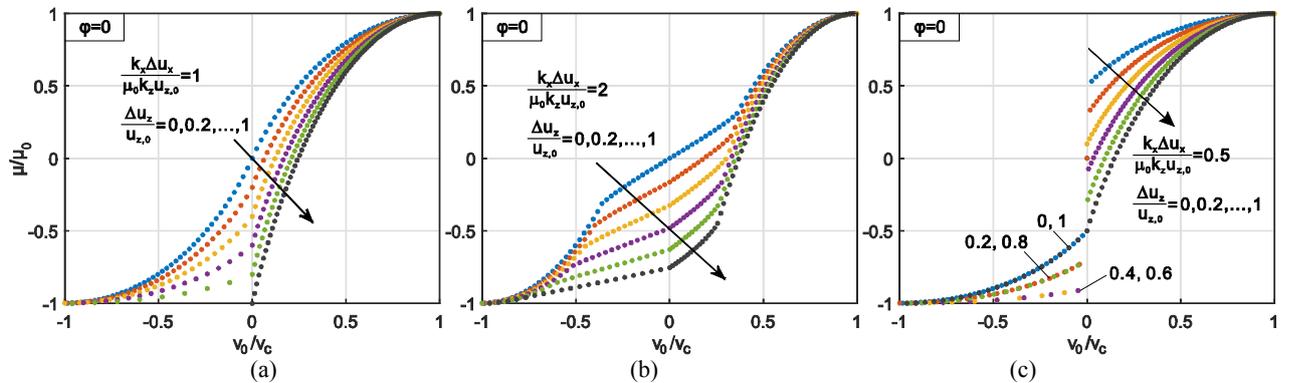

**Fig. 4** A few examples of dependences of the coefficient of friction on velocity with phase difference $\varphi = 0$, $\frac{\Delta u_z}{\Delta u_{z,0}} = 0,...,1$ and (a) $\frac{k_x \Delta u_x}{\mu_0 k_z u_{z,0}} = 1$; (b) $\frac{k_x \Delta u_x}{\mu_0 k_z u_{z,0}} = 2$; (c) $\frac{k_x \Delta u_x}{\mu_0 k_z u_{z,0}} = 0.5$.

Note that the critical velocity $v_c$ is different for positive and negative sliding velocities (see Eq. (8)). The two sides of Fig. 4. are normalized to these two different critical velocities, resulting in a visible "kink" at $v_0/v_c = 0$ in many of the curves.

In Fig. 5 to Fig. 7, on the other hand, we use dimensional (non-normalized) velocity to get a more clear physical picture. These figures show the dependence of the coefficient of friction on the sliding velocity for different values of $k_x \Delta u_x / (\mu_0 k_z \Delta u_z)$ and different phase shifts while $\Delta u_z / u_{z,0}$ is fixed.

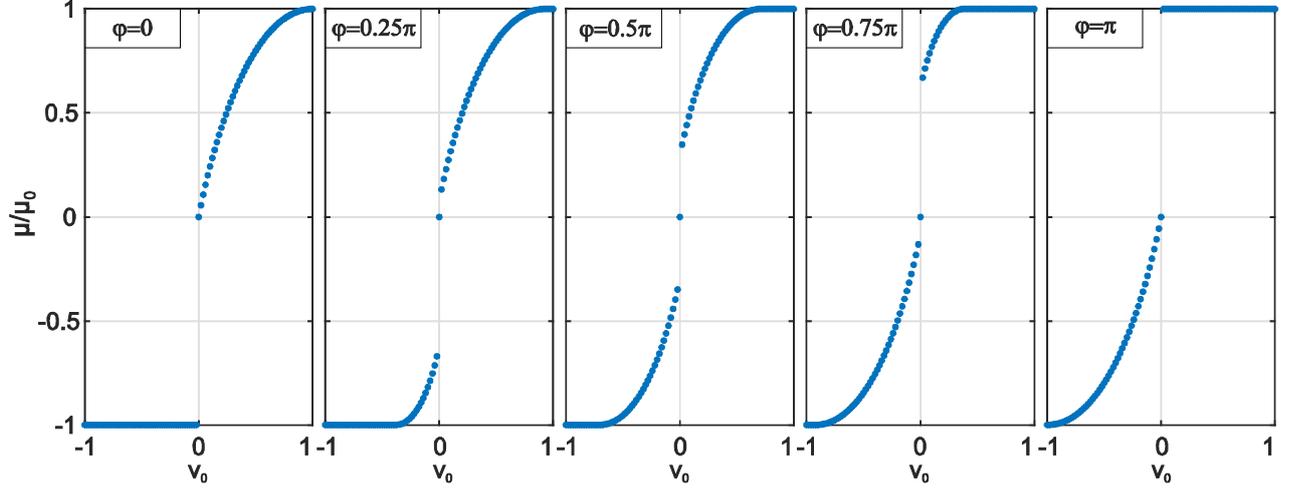

**Fig. 5** Velocity-dependences of the coefficient of friction for different phase shifts $\varphi$ for the case $\dfrac{k_x \Delta u_x}{\mu_0 k_z \Delta u_z} = 1$ and $\dfrac{\Delta u_z}{u_{z,0}} = 0.5$.

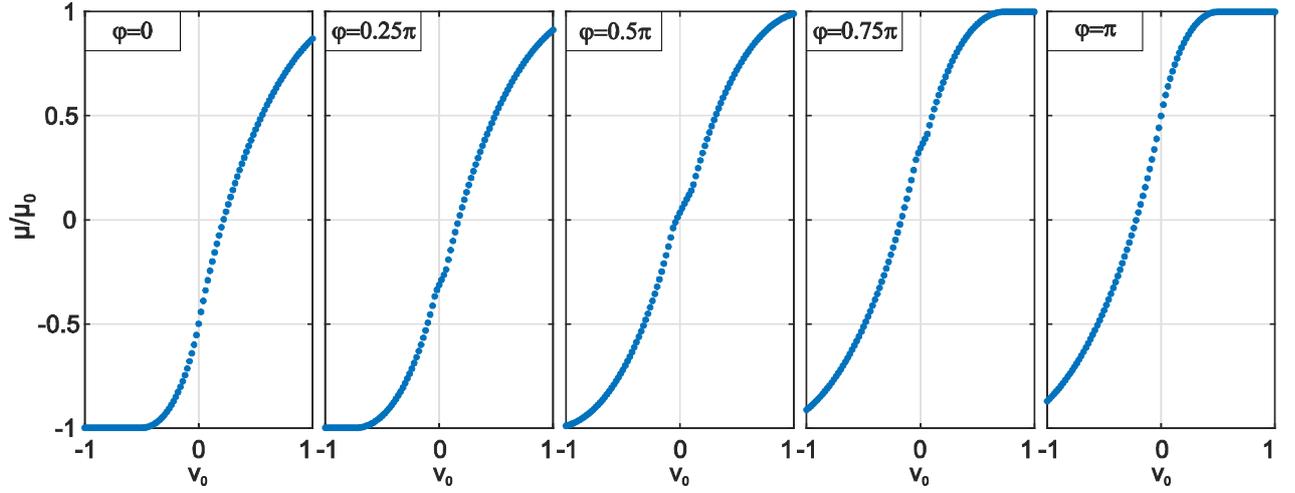

**Fig. 6** Velocity-dependences of the coefficient of friction for the case $\dfrac{k_x \Delta u_x}{\mu_0 k_z \Delta u_z} = 2$ and $\dfrac{\Delta u_z}{u_{z,0}} = 0.5$.

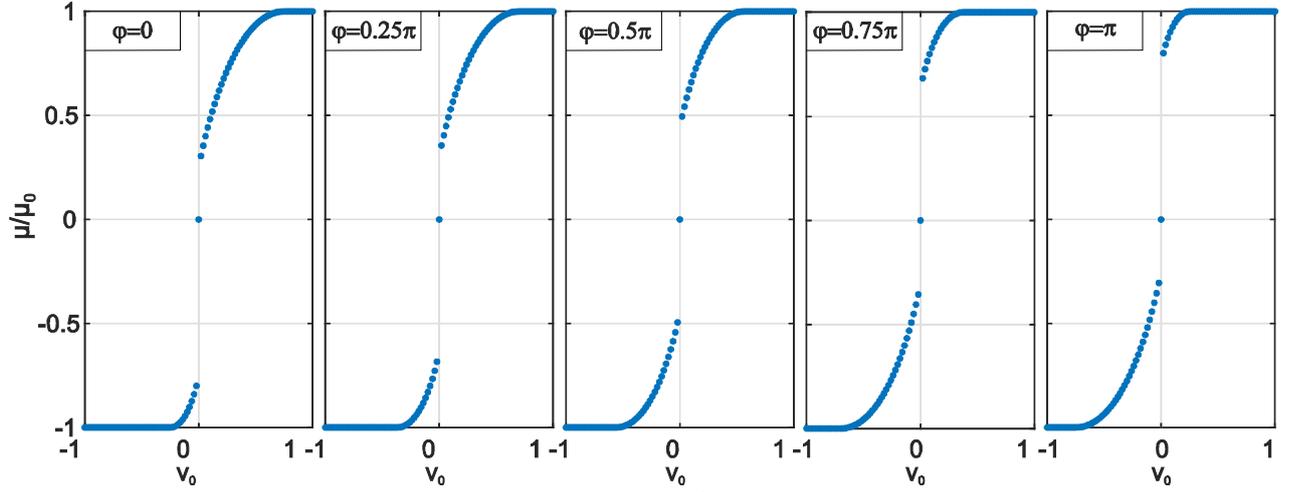

**Fig. 7** Velocity-dependences of the coefficient of friction for the case $\dfrac{k_x \Delta u_x}{\mu_0 k_z \Delta u_z} = 0.5$ and $\dfrac{\Delta u_z}{u_{z,0}} = 0.5$.

In Fig. 5, corresponding to $\dfrac{k_x \Delta u_x}{\mu_0 k_z \Delta u_z} = 1$, for $\varphi = \dfrac{\pi}{2}$ the dependence is symmetrical with respect of change of the sign of the velocity. For other phase shifts the dependencies are asymmetrical and become extremely asymmetrical at $\varphi = 0$ and $\varphi = \pi$. The first and the last subplots correspond to the particular cases $v_c = 0$ for $\dfrac{k_x \Delta u_x}{\mu_0 k_z \Delta u_z} = 1$, and $\varphi = \pi$ at positive velocity and $\varphi = 0$ at negative velocity (see Eqs. (10) and (18)).

The dependences of the coefficient of friction on the sliding velocity presented in Figs. 2 to 7 can be classified in three qualitatively different categories:

I. *Active control of friction.*
If the dependence is symmetrical with respect to reversing the sign of velocity, we have a classical "law of friction". The external oscillation influences the static force of friction and the velocity-dependence but does not change the basic character of the force as a dissipative force which is always directed opposite to the velocity. This case is found (always) with single-mode oscillations (Fig. 2 and Fig. 3) as well as with dual-mode oscillations when $\varphi = 0.5\pi$ (Fig. 5 to Fig. 7).

II. *Dynamic ratchet.*
Into the second class fall dependences that are asymmetrical with respect to change of the sign of velocity but remain "dissipative" (thus, the force of friction is still directed opposite to the velocity). The most extreme case in this category is represented by the first and the last subplots in Fig. 5. In these extreme cases, the static coefficient of friction for backward movement is $\mu_0$ and zero for forward movement at $\varphi = 0$ and vice versa at $\varphi = \pi$. This means that if the *substrate* is subjected to a low-frequency tangential force, it will move forth in the positive half-period and will stick in the negative half-period, thus resembling the action of a mechanical ratchet. We therefore call this class a "dynamic ratchet".

III. *Drive or actuator.*
Finally, we have the cases where the "law of friction" is not only asymmetric but "active", in that the direction of the average tangential force is *opposite* to the direction of movement at small velocities. This is functionally equivalent to a vibrational drive or actuator. This case is represented by all curves in Fig. 4a and Fig.4b with the exception of the upper-most curve, which corresponds to a purely horizontal oscillation. The same situation can be found in Fig. 6 for $\varphi \neq \pi/2$.

## 4 Conclusion

We considered bi-modal control of friction by a superposition of normal and tangential (in the direction of motion) oscillations. In the presence of oscillations in both directions, the dependence of the macroscopic coefficient of friction (which is here formally defined as the normalized tangential force and can assume both positive and negative values) on the macroscopic sliding velocity becomes asymmetric in the general case. While the asymmetry as such is understandable from general considerations (see e.g. [17]), the detailed form of the laws of friction and their classification seems to be non-trivial and has not been described earlier. In particular, apart from known effects of active control of friction on one hand and oscillation induced actuation on the other hand, we predict a third, intermediate type of behavior which we call "dynamic ratchet". Dynamic ratchets are realized for values of the governing parameter $\frac{k_x \Delta u_x}{\mu_0 k_z \Delta u_z}$ smaller than 1 while drives result with $\frac{k_x \Delta u_x}{\mu_0 k_z \Delta u_z} > 1$.

Finally, let us note that the one-spring model is not an essential assumption for the described qualitative behavior. As any contact can be mapped to a contact with a one-dimensional elastic foundation [18],[19], this analysis can be easily generalized.

## 5 Acknowledgment

The authors thank V.L. Popov for valuable discussions. This work was partially supported by the German Academic Exchange Service (DAAD).